\documentclass[12pt]{article}

\usepackage{graphicx}
\usepackage{amssymb,amsfonts,amsmath,picins,multirow}
\usepackage{color} 

\newcommand{\affil}[1]{${}^{#1}$}
\newcommand{\contributor}[1]{}
\newcommand{\dropcap}[1]{#1}
\newcommand{\keywords}[1]{}
\newenvironment{acknowledgments}{\section*{Acknowledgements}}{}
\newenvironment{article}{}{}

\begin{document}

\title{Uncovering space-independent communities in spatial networks\thanks{Published as PNAS, 2011, \textbf{108}, 7663-7668. \texttt{arXiv:1012.3409}. Original title: ``Beyond Space for Spatial networks''.}}


\author{Paul Expert\affil{1}\affil{2},
 Tim S. Evans\affil{3},
 Vincent D. Blondel\affil{4},
 Renaud Lambiotte\affil{1,5}
 \\[0.5cm] \affil{1} Institute for Mathematical Sciences, Imperial College London,  \\ London SW7 2PG, U.K.
 \\ \affil{2} Blackett Laboratory, Prince Consort Road, Imperial College London, \\ London SW7 2AZ, U.K.
 \\ \affil{3} Theoretical Physics, Imperial College London, \\ London SW7 2AZ, U.K.
 \\ \affil{4} Massachusetts Institute of Technology, \\ Laboratory for Information and Decision
Systems, \\77 Massachusetts Avenue, Cambridge, MA 02139, USA
 \\ \affil{5} Naxys, Facult{\' e}s Universitaires Notre-Dame de la Paix, \\B-5000 Namur, Belgium
}

\date{7th March 2011}

\contributor{Submitted to Proceedings of the National Academy of Sciences
of the United States of America}

\maketitle

\begin{article}

\begin{abstract} Many complex systems are organized in the form of a network embedded in space. Important examples include the physical Internet infrastucture, road networks, flight connections, brain functional networks and social networks.  The effect of space on network topology has recently come under the spotlight because of the emergence of pervasive technologies based on geo-localization, which constantly fill databases with people's movements and thus reveal their trajectories and spatial behaviour.  Extracting patterns and regularities from the resulting massive amount of human mobility data requires the development of appropriate tools for uncovering information in spatially-embedded networks. In contrast with most works that tend to apply standard network metrics to any type of network, we argue in this paper for a careful treatment of the constraints imposed by space on network topology. In particular, we focus on the problem of community detection and propose a modularity function adapted to spatial networks. We show that it is possible to factor out the effect of space in order to reveal more clearly  hidden structural similarities between the nodes. Methods are tested on a large mobile phone network and computer-generated benchmarks where the effect of space has been incorporated.
\end{abstract}

\keywords{complex networks | social systems | spatial networks}


\dropcap{U}nderstanding the principles driving the organization of complex networks is crucial for a broad range of fields including information and social sciences, economics, biology and neuroscience \cite{review2}. In networks where nodes occupy positions in an Euclidian space, spatial constraints may have a strong effect on their connectivity patterns \cite{bart_review}. Edges may either be spatially-embedded, such as in roads or railway lines in transportation networks or cables in a power grid, or abstract entities, such as friendship relations in online and offline social networks or functional connectivity in brain networks. In either case, space plays a crucial role by affecting, directly or indirectly, network connectivity and making its architecture radically different from that of random networks \cite{gast1}. A crucial difference stems from the cost associated to long-distance links \cite{cost1,Amaral00,Amaral04,cost2,cost2b,cost3,cost4,liben,cost5} which restricts the existence of hubs, i.e.\ high degree nodes, and thus the observation of fat-tailed degree distributions in spatial networks.

From a modeling viewpoint, gravity models \cite{gr1,gr3,gr5} have long been used to model flows in spatial networks. These models focus on the intensity of interaction between locations $i$ and $j$ separated by a certain physical distance $d_{ij}$. It has been shown for systems as diverse as the International Trade Market \cite{gravity1}, human migration \cite{gravity1b}, traffic flows \cite{gravity2} or mobile communication between cities \cite{mobile,mobile2} that the volume of interaction between distant locations is successfully modeled by
\begin{equation}
\label{pop}
T_{ij}=N_i N_j f(d_{ij}),
\end{equation}
where $N_i$ measures the importance of location $i$, e.g. its population, and the deterrence function $f$ describes the influence of space.
Equation (\ref{pop}) emphasizes that the number of interactions between two locations is proportional to the number of possible contacts $N_i N_j$ and that it varies with geographic distance, because of financial or temporal cost. In many socio-economic systems, $f$ is well fitted by a power law $\sim d_{ij}^{-\alpha}$ reminiscent of Newton's law of gravity, with population playing the role of a mass.

Whereas a broad range of models have been specifically developed for spatial networks \cite{kleinberg,bart,Manna,gast2,wong}, dedicated tools for uncovering useful information from their topology are poorly developed. When analyzing spatial networks, authors tend to use network metrics where the spatial arrangement of the nodes is ignored, thus disregarding that useful measures for non-spatial networks might yield irrelevant or trivial results for spatial ones. Important examples are the clustering coefficient, as spatial networks are often spatially clustered by nature, and degree distribution, where high degree nodes are suppressed by long distance costs. This observation underlines the need for appropriate metrics for the analysis and modeling of networks where spatial constraints play an important role \cite{kosmidis,mascolo,colizza}.

This need is particularly apparent in the context of community detection. The detection of communities (modules or clusters) is a difficult task which is important to many fields, and it has attracted much attention in the last few years \cite{GN,danon,newman_modul_PNAS,Amaral05a,sune,porter,fort}. In a nutshell, modules are defined as sub-networks that are locally dense even though the network as a whole is sparse. Community detection is a central tool of network theory because revealing intermediate scales of network organization provides the means to draw readable maps of the network and to uncover hidden functional relations between nodes  \cite{Amaral05a}. In the case of spatial networks, important practical applications include: i) the design of efficient national, economical or administrative borders based on human mobility or economical interactions instead of historical or ad-hoc reasons \cite{brockmann10,brockmann10b,gauthier,ratti}; ii) the modeling of historical or pre-historical interactions based on limited archaeological evidence \cite{RW87,KER08}; iii) the identification of functionally related brain regions and of principles leading to global integration and functional segregation \cite{bassett,frontiers}.

In practice, the current state-of-the-art for finding modules in spatial networks \cite{Guimera,onnela2010} is to optimize the standard Newman-Girvan modularity which, as we argue below, overlooks the spatial nature of the system. In most cases, this scheme produces communities which are strongly determined by geographical factors and provide poor information about the underlying forces shaping the network. For instance, social and transportation networks are typically dominated by low cost short-ranged interactions leading to modules which are compact in physical space. As a result, modularity optimization is blind to spatial anomalies and fails to uncover modules determined by factors other than mere physical proximity. This point brings us to the central question of our work: In spatial networks, how can one detect patterns that are not due to space? In other words, are observed patterns only due to the effect of spatial distance, because of gravity-like forces, or do other forces come into play? If that is the case, can one go beyond a standard network methodology in order to uncover significant information from spatial networks?

\section{Social Networks and Space}

In order to illustrate these concepts and to clarify the goal of this paper, let us elaborate on social networks, where the dichotomy between network and space has been studied for decades. On the one hand, research has attempted to explain the organization of social networks purely in terms of the structural position of the nodes. Structural mechanisms underpinning the existence of social interactions include triadic closure \cite{davis70}, link reciprocity \cite{gouldner60} and reinforcement \cite{podolny94}. On the other hand, research has identified ordering principles that explain edge creation in terms of non-structural attributes, mainly homophily \cite{mcpherson01} and focus constraint \cite{feld81}. Homophily states that similarity, e.g.\ in terms of status or interests, fosters connection \cite{mcpherson01}, as similar people tend to select each other, communicate more frequently and develop stronger social interactions \cite{kossinets06}.  The second ordering principle is focus constraint \cite{feld81}, which refers to the idea that social relations depend on opportunities for social contact. A dominant factor for focus constraint is geographic proximity, which offers opportunities for face-to-face interaction and encounters between individuals \cite{owen04, crandall}. Focus constraint thus depends indirectly on distance through its dependence on transportation networks which themselves typically exhibit a gravity law.

Although homophily and focus constraint are different mechanisms, they are often inter-related, because frequent contacts drive groups towards uniformity, through social influence, and that alike individuals tend to live in the same neighborhoods \cite{schelling}. Moreover, both aspects can be seen as originating from proximity in a high-dimensional social space, which summarizes people's interests and characteristics, i.e.\ nodes have a tendency to connect with neighbouring nodes in social space \cite{watts}. When uncovering modules of strongly connected nodes in complex networks, one deals with an extremely intricate situation where structural and non-structural effects, including homophily and focus constraint, are mingled. Modules uncovered by community detection are thus underpinned by an uncontrolled mixture of possibly antagonistic forces, from which few conclusions can be drawn \cite{pietro}. Our aim is the following: when the spatial positions of the nodes are known, as more and more often is the case, is it possible to take out the effect of space in order to identify more clearly homophilious effects and thus hidden structural or cultural similarities.

\section{Modularity and Space}
Let us now introduce the notations and formalize the problem of community detection.
In the following, we focus on weighted, undirected networks characterized by their adjacency matrix $A$. By definition, $A$ is symmetric and $A_{ij}$ is the weight of the link between $i$ and $j$. The strength of node $i$ is defined as $k_i = \sum_{j} A_{ij}$; $m = \sum_{i,j} A_{ij}/2$ is the total weight in the network. The distance between nodes $i$ and $j$ is denoted by $d_{ij}$. From now on, by distance, we mean Euclidian distance between nodes when measured on the embedding space, and not network distance, which is the number of edges traversed along the shortest path from one vertex to another. As discussed above, the nature of space and its associated distance may be abstract, i.e.\ affinity in a social network, or physical, i.e.\ geographical distance between cities.

The fundamental idea behind most community detection methods is to partition the nodes of the network into modules. Contrary to standard graph partitioning algorithms, the detection of communities is performed without a priori specifying  the number of modules nor their size, and aims at uncovering in an automated way the meso-scale organization of the network \cite{newman_modul_PNAS}. Behind most community detection methods, there is a mathematical definition measuring the quality of a partition. The widely-used modularity \cite{NG} of a partition $\mathcal P$ measures if links are more abundant within communities than would be expected on the basis of chance, namely
\begin{align}
Q &= \mbox{(fraction of links within communities)} \nonumber\\
  &\qquad{} - \mbox{(expected fraction of such links)}
\label{eq:modDeff}
\end{align}
In a mathematical expression, modularity reads
\begin{equation}
\label{modularity}
Q = {1\over2m} \sum_{C \in \mathcal{P}} \sum_{i,j \in C} \biggl[ A_{ij} - P_{ij} \biggr],
\end{equation}
where  $i,j \in C$ is a summation over pairs of nodes $i$ and $j$ belonging to the same community $C$ of $\mathcal P$ and therefore counts links between nodes within the same community.

What is meant by chance, i.e.\ the null hypothesis, is an extra ingredient in the definition \cite{opt2} and is embodied by the matrix $P_{ij}$. $P_{ij}$ is the expected weight of a link between nodes $i$ and $j$  over an ensemble of random networks with certain constraints. These constraints correspond to known information about the network organization, i.e., its total number of links and nodes, which has to be taken into account when assessing the relevance of an observed topological feature. In general, if $A_{ij}$ is symmetric, $P_{ij}$ is also chosen to be symmetric and one also imposes that the total weight is conserved\footnote{This constraint can be relaxed in order to change the characteristic size of the network and thus to tune the resolution at which communities are uncovered  \cite{reichardt} (see SI)}, i.e.\ $\sum_{ij} A_{ij} =\sum_{ij} P_{ij}=2m$. Beyond these basic considerations, different null models can be constructed depending on the network under consideration \cite{Amaral07_1,gen1,gen2}. The most popular choice, proposed by Newman and Girvan (NG) \cite{NG} is
\begin{eqnarray}
P^{\rm NG}_{ij}= k_i k_j/2m, &  \text{ then $Q =  Q_{\rm NG}$.}
\end{eqnarray}
where randomized networks preserve the strength of each node. Constraining the node strengths goes along the view that the network is well-mixed, in the sense that any node can be connected to any node and that only connectivity matters. In that case, node strength is a good proxy for the probability of a link to arrive on a certain node. Different types of heuristics can be developed in order to approximate the optimal value of the corresponding NG modularity \cite{opt2,opt1,opt3,spectmas}. These methods have been shown to produce useful and relevant partitions in a broad class of systems \cite{newman_modul_PNAS}, even if modularity suffers from limitations such as resolution limit \cite{FB} and a possible high degeneracy of its landscape  \cite{Amaral07_2,Good}.

The NG null-model only uses the basic structural information encoded in the adjacency matrix. Therefore, it is appropriate when no additional information on the nodes is available but not when additional constraints are known.  In networks where distance strongly affects the probability for two nodes to be connected, a natural choice for the null model is inspired by the  afore-mentioned gravity models
\begin{eqnarray}
\label{spatial}
    P^{\rm Spa}_{ij}=N_{i}N_{j} f(d_{ij})
\end{eqnarray}
where $N_{i}$ is, as in (\ref{pop}), a notion of importance of node $i$ and where the deterrence function
\begin{eqnarray}
f(d) = \frac{\sum_{i,j|d_{ij}=d}A_{ij}}{\sum_{i,j|d_{ij}=d}N_{i}N_{j}},
\end{eqnarray}
is the weighted average of the probability $A_{ij}/(N_i N_j)$ for a link to exit at distance $d$. It is thus directly measured from the data\footnote{In practice, when analyzing empirical data, the distance between 2 cities is binned such as to smoothen $f(d)$. The dependence of our results on bin size is explored in SI.} and not fitted by a determined functional dependence, as is often the case \cite{gr5}. By construction, the total weight of the network is conserved as required. Depending on the system under scrutiny, $N_i$ may be the number of inhabitants in a city or the degree of a node when it corresponds to a single person in a social network. It is worth mentioning that in the latter case and if the embedding in space does not play a role, i.e.\ where $f(d)$ is flat, the standard NG model is exactly recovered (see SI).

From now on, let us denote by $Q_{\rm Spa}$ the version of modularity (\ref{modularity}) whose null model $P^{\rm Spa}_{ij}$ is given by (\ref{spatial}). $Q_{\rm Spa}$ incorporates non-structural information about the nodes, i.e.\ their position in physical space. By definition, it favours communities made of nodes $i$ and $j$ such that $A_{ij}-P^{\rm Spa}_{ij}$ is large, i.e. pairs of nodes which are more connected than expected for that distance. Compared to $Q_{\rm NG}$, $Q_{\rm Spa}$ tends to give larger contributions to distant nodes and its optimization is expected to uncover modules driven by non-spatial factors.

\section{Numerical validation}

\subsection{Belgian mobile phone data}
In order to compare the partitions obtained by optimizing $Q_{\rm NG}$ and $Q_{\rm Spa}$, let us first focus on a Belgian mobile phone network made of its $571$ communes (the 19 communes forming Brussels are merged into one) and of the symmetrized number of calls $\{A_{ij}\}_{i,j=1}^{571}$ between them during a time period of $6$ months (see \cite{gauthier} for a more detailed description of the data). This network is aggregated from the customer-customer communication network of a large mobile phone provider by using the billing address associated to each customer. The number of customers in each commune $i$ is given by $N_{i}$. This network provides an ideal test for our method because of the importance of non-spatial factors driving mobile phone communication, namely the existence of two linguistic communities in Belgium\footnote{There also exist a German-speaking community made of only 0.73\% of the national population}: a Flemish community and a French community mainly concentrated in the North and the South of the country respectively. As reported in \cite{gauthier}, when the weights between communes are given by the average duration of communication between people in $i$ and $j$, a standard NG modularity optimisation recovers a bi-partition that closely follows the linguistic border.

\begin{figure}
    \centering
    \centerline{\includegraphics[width=8cm,angle=0]{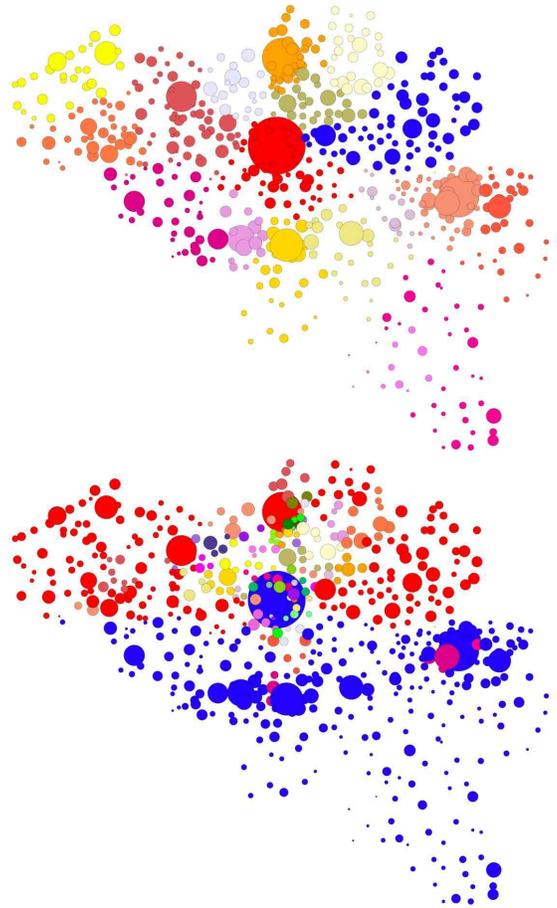}}
    \caption{ Decomposition of a Belgian mobile phone network into communities (see main text). Each node represents a commune and its size is proportional to its number of clients $N_i$. (Top figure) Partition into 18 communities found by optimizing NG modularity. (Bottom figure) Partition into 31 communities found by optimizing  Spa modularity.\label{mobile}}
\end{figure}

Both versions of modularity are optimized using the spectral method described in \cite{spectmas}. Visualization of the results are shown in Fig.\ref{mobile}. The NG modularity uncovers 18 spatially compact modules, similar to those observed in other spatially extended networks and mainly determined by short-range interactions between communes. Although this partition coincides with the linguistic separation of the country \cite{gauthier}, the unaware would not discover the existence of two linguistic communities only from Fig.\ref{mobile}. The spatial modularity uncovers a strikingly different type of structure: an almost perfect bipartition of the country where the two largest communities account for about 75\% of all communes (see SI for more details) and nicely reproduce the linguistic separation of the country. Moreover, Brussels is assigned to the French community, in agreement with the fact that $\approx 80\%$ of its population is French speaking, and despite the fact that it is spatially located in Flanders. The remaining smaller communities (not bigger than 10 communes each) originate from the constraints imposed by a hard partitioning, which is blind to overlapping communities and might thus misclassify Flemish communes strongly interacting with Brussels and communes that have mixed language populations. A similar bipartition is found by considering only the signs of the dominant eigenvector of the modularity matrix (see SI).

\subsection{Statistical tests}
The values for the optimal modularities can be found in Table \ref{tableZ}. It is important to stress that a direct comparison of $Q_{\rm NG}$ and $Q_{\rm Spa}$ is meaningless, as modularity is a way to compare different partitions of the same graph and so its absolute value is inconsequential. Moreover, the value of modularity is expected to be lower when its null model is closer to the real structure of the data, as it is the case for $Q_{\rm Spa}$. In order to assess the significance of the uncovered partitions, one needs instead to resort to statistical tests by comparing modularity with that of an ensemble of random networks \cite{opt1}.

\begin{table}
\begin{center}
    \begin{tabular}{c|c|c c c}
    \multicolumn{2}{c|}{} & $Q^{\rm norm}_{\rm obs}$ & $\langle Q^{\rm norm}_{\rm rand}\rangle$ & $z$ score\\
    \hline
    \multirow{2}{*}{Weights}& Spatial & 0.0881 & 0.0049 & 803 \\
    & NG & 0.7961 & 0.7059 & 55 \\
    \hline
    Positions & Spatial & 0.0881 & 0.2383 & -90 \\
    \hline
    \end{tabular}
\end{center}
    \caption{$z$-scores for the modularity measured on the original data-set compared to the modularity values measured on the randomized data.\label{tableZ}}
\end{table}

Two types of random networks are constructed: i) Networks where weights are randomized. Starting from the empirical $f(d)$, we generated weights between two communes $i$ and $j$ according to a binomial of mean $\rho N_{i}N_{j}f(d_{ij})$. In the following, we chose $\rho=1$, thus conserving (up to some fluctuations) the total number of calls in the system and the spatial dependence between nodes. Let us keep in mind that $\rho$ allows to tune the importance of finite size fluctuations and that $A_{ij}/\rho=N_{i}N_{j}f(d_{ij}))$ in the limit $\rho \rightarrow \infty$.  ii) Networks where the geographical position of the nodes is randomized while leaving the weights unchanged. This second ensemble of random networks is radically different from the first one because it keeps the topology of the network unaffected and only randomizes node attributes. Since NG does not make use of geographical information, it is unaffected by this reshuffling. By construction, the effect on Spa is to make space less important by changing the function $f(d)$, thus leading to an expression closer to NG (see SI). For each type of randomization we produce $N=100$ networks and optimize their modularities $Q_{\rm NG}$ and $Q_{\rm Spa}$.

The significance of the partitions found in the original data is first evaluated by comparing their modularity with that of the randomized data through a $z$-score \cite{opt1}, defined as
\begin{equation}
z = \frac{Q - \langle Q\rangle_{\rm random}}{\sigma},
\end{equation}
where $\sigma$ is the standard deviation across $100$ realizations. Results are summarized in Table \ref{tableZ} and clearly show that the original data is significantly more modular than networks where the weights are randomized. The $z$-score is an order of magnitude larger for the spatial modularity. For the spatial randomization, in contrast, the $z$-score is negative, which reflects the fact that useful information is lost by randomizing node positions and that the resulting randomized null-model is further away from reality than the original.

As a next step, we focus on the variability across the uncovered partitions. This is done by using normalized variation of information (VI) \cite{varinfo}, which is a measure of the distance between partitions. VI is equal to 0 only when two partitions are identical and is between 0 and 1 otherwise. Results are summarized in Table \ref{TableRnd} where we observe that partitions obtained from NG and Spatial are genuinely different. In the case of weight randomization, the important point is that VI between partitions uncovered in random networks is much smaller for NG ($0.09$) than for Spa ($0.58$), thus indicating that very similar partitions are found by NG across random networks, i.e.\ only due to spatial interactions between communes. Another interesting point is the high similarity between partitions found by NG in the original data and by Spa in the spatially randomized networks, as their VI is found to be equal to $0.16$, in agreement with the fact that Spa becomes similar to NG when space is irrelevant \footnote{It is important to stress that the spatial randomization does not entirely remove the effect of space on network connectivity because self-loops, i.e. intra-commune links, are preserved.}. This observation is confirmed by the similar values of VI between the partitions found by NG and Spa in the original data, as shown in Fig.\ref{mobile}, i.e. $0.38$, and between partitions found by Spa in the original data and in the spatially randomized data ($0.35$ in Table \ref{TableRnd}).

\begin{table}
    \begin{center}
    \begin{tabular}{c|c|c|c}
    \multicolumn{2}{c|}{} & Orig-Rand & Rand-Rand\\
    \hline
    \multirow{2}{*}{Weights}& Spatial & $0.54\pm0.02$ & $0.58\pm0.02$\\
    & NG &  $0.23\pm0.02$ & $0.09\pm0.05$\\
    \hline
    Positions& Spatial & $0.35\pm0.02$ & $0.07\pm0.04$\\
    \hline
    \end{tabular}
    \end{center}
    \caption{Average VI measured between the partition found on the original data-set and the randomized ones (Orig-Rand) and the average VI among the randomized data-set (Rand-Rand) for both null-models and randomization procedures. \label{TableRnd}}
\end{table}

\subsection{Gravity Model Benchmark}
In order to test the validity of our method in a controlled setting, let us now focus on computer-generated benchmarks for spatial, modular networks. The underlying idea is to build spatially-embedded random networks where the probability for two nodes to be connected depends on their distance, as observed in real-world examples, and on the community to which they are assigned. We implement benchmarks in the simplest way by throwing $100$ nodes at random in a two dimensional square of dimension $100 \times 100$ and by randomly assigning them into two communities of $50$ nodes. Contrary to the previous example, where nodes (communes) could have different sizes, we assume that all nodes have the same size. The probability that a link exists between nodes $i$ and $j$ has the form
\begin{eqnarray}
 p_{ij}=\frac{\lambda(c_i,c_j)}{Z d_{ij}},
\end{eqnarray}
\noindent where $c_i$ is the community of node $i$.  The function $\lambda(c_i,c_j)$ determines the community linkage. By definition, it is equal to 1 if $c_i=c_j$ and $\lambda_{\rm different}$ otherwise.  When $\lambda_{\rm different}=0$, only nodes in the same community are connected, while no distinct communities are present when $\lambda_{\rm different}=1$. A normalization constant, $Z$, ensures that $\sum_{i>j} p_{ij}=1$. These networks, directly inspired by gravity models, are built by placing $L=\rho N(N-1)/2$ links with probability $p_{ij}$, where $\rho \geq 0$ determines the density of links in the network. Multiple links are allowed and interpreted as weights. The parameter $\rho$ controls finite-size fluctuations around the expected number of edges $L p_{ij}$.

\begin{figure}
    \centering
   \centerline{ \includegraphics[width=8cm,angle=0]{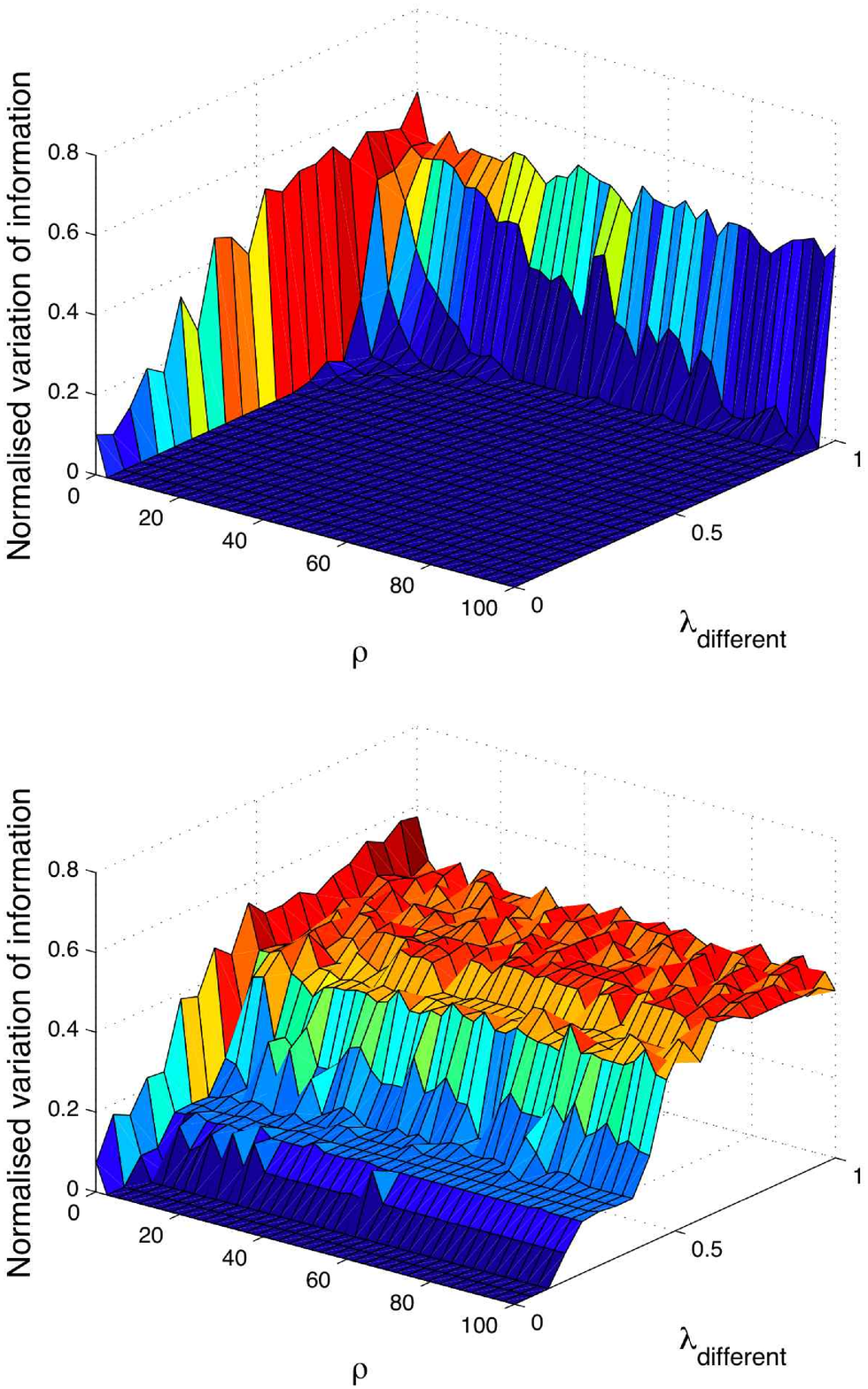}}
    \caption{Variation of information over the $(\lambda_{\rm different},\rho)$ parameter space for Spa (top figure) and NG (bottom figure) when tested on the spatial benchmark. Spa is able to recover the correct communities over a wide range of parameters' values, while NG fails to find the correct communites almost as soon as the interaction $\lambda_{\rm different}$ is turned on.\label{VIBench}}
\end{figure}

In order to compare the efficiency of $Q_{\rm Spa}$ and $Q_{\rm NG}$, we generated one realization of the random model for different values of $\lambda_{\rm different}\in[0,1]$ and $\rho\in[0.01,100]$, and optimized their modularity. As a measure of the quality of the uncovered partitions, we compared them with the known bipartition of the network by using normalized VI. Our simulations show that $Q_{\rm Spa}$ outperforms $Q_{\rm NG}$ and that the improvement becomes larger and larger as the density of links is increased (see Fig.\ref{VIBench}). In the limit $\rho \rightarrow \infty$, where fluctuations become negligible, our simulations show that Spa perfectly identifies the correct communities for any $\lambda_{\rm different} <1$ while NG fails even for small values of $\lambda_{\rm different}$. It is also interesting to note that results presented in Fig.\ref{VIBench} are obtained for single realizations of the random networks, i.e.\ as when dealing with empirical data sets one does not analyze an ensemble of networks, and yet the precision of Spa is significantly better than that of NG (results smoothed by averaging over several realizations are presented in SI).

\section{Discussion}

Despite the increasing availability of affordable long distance travel and new communication media, the ``death of distance''   \cite{cairncross97} has been greatly exaggerated \cite{liben,spiky}. Furthermore, the emergence of new technologies entangling physical and virtual worlds has stimulated new research and produced new applications for social and human mobility networks embedded in space \cite{mobility}. This importance of space is not limited to social networks as a broad range of economical and biological are also spatially embedded, with strong consequences on their topological organization.
The main purpose of this paper has been to find new ways to uncover significant patterns in spatial networks. To do so, we have taken advantage of the flexibility of a quantity called modularity defined for community detection. Modularity incorporates a null model which represents what is expected by chance, namely the expected probability that two nodes are connected. Unlike the standard null model, we incorporate non-structural attributes into our null model and use this as a comparison with empirical data. By doing so, we construct null models which portray more closely the network under scrutiny and provide the means to exploit known attributes, e.g.\ spatial location, in order to uncover unknown ones, e.g.\ homophilious relations.

We believe that our general framework is suitable for a wide range of networks and that it opens avenues of quantitative exploration of spatially distributed systems.
Interesting lines of research include the development of more general null models, for instance by interlacing structural and non-structural information, and the detection of hierarchies in spatial networks either by tuning the resolution of modularity \cite{reichardt}  or by looking for local maxima of the modularity landscape \cite{Amaral07_2} (see SI for more details).
Moreover, our methodology is not limited to situations where distance is measured in physical space as it may be applied whenever one can use node attributes to define a separation between nodes.  For instance in many social networks age may be a dominant factor, yet by building a null model on the age difference between actors, other types of relationships may be revealed for little extra computational effort. A further advantage is that by incorporating relevant information, a partitioning approach can be applied even if modules are pervasively overlapping \cite{ABL09,EL10}.

\begin{acknowledgments}
We would like to thank K. Christensen, M. Gastner and H.J.\ Jensen for fruitful discussions and N. Zachariou for proof-reading.
R.L. acknowledges support from the UK EPSRC. This work was conducted within the framework of COST Action MP0801 Physics of Competition and Conflicts.
\end{acknowledgments}


\end{article}
\appendix
\section*{Supplementary Information}

This supplement to the paper ``Uncovering space-independent communities in spatial networks'' contains detailed information on relations between the spatial null model and the standard NG null model, multi-scale modularity, data-sets, effect of bin size on results, optimal bi-partitions of the mobile phone networks, randomizations used to assess the significance of the results.

\section{Spatial vs Newman-Girvan null model}

As discussed in the main text, the null model $P_{ij}$ is a crucial ingredient of modularity defined as
\begin{equation}
Q = {1\over2m} \sum_{C \in \mathcal{P}} \sum_{i,j \in C} \biggl[ A_{ij} - P_{ij} \biggr].
\end{equation}
The most standard choice is
\begin{eqnarray}
P^{\rm NG}_{ij}= k_i k_j/2m
\end{eqnarray}
where the probability for 2 nodes to be connected is proportional to their degree. Our spatial null model incorporates non-structural ingredients, namely a dependence on the physical distance $d_{ij}$ between 2 nodes
\begin{eqnarray}
P^{\rm Spa}_{ij}=N_{i}N_{j} f(d_{ij})
\end{eqnarray}
where $N_{i}$ is a measure of the importance of node $i$ and where the deterrence function
\begin{eqnarray}
f(d) = \frac{\sum_{i,j|d_{ij}=d} A_{ij}}{\sum_{i,j|d_{ij}=d}N_{i}N_{j}}
\end{eqnarray}
is measured from the empirical data. This expression directly comes from the constraint
\begin{eqnarray}
\sum_{i,j|d_{ij}=d} P^{\rm Spa}_{ij}=\sum_{i,j|d_{ij}=d} A_{ij}
\end{eqnarray}
that the total weights between nodes at a certain distance is preserved. When analysing the mobile phone network, we have taken $N_i$ as the number of clients in commune $i$, in analogy with simple versions of gravity models. In that case, the above expression for $f(d)$ is a weighted average of the probability $A_{ij}/(N_{i}N_{j})$ to have a call between clients in $i$ and in $j$.

An interesting choice for $N_i$ is to chose the degree itself, i.e. $N_i=k_i$ such that the set of null models
\begin{eqnarray}
P^{\rm Spa}_{ij}=k_i k_j f(d_{ij})
\end{eqnarray}
includes $P^{\rm NG}_{ij}$. Indeed, if $f(d)$ does not depend on $d$, one finds $f(d)=1/2m$ and $P^{\rm Spa}_{ij}$ reduces to $P^{\rm NG}_{ij}$.

Finally, we would like to briefly introduce a possible further generalization of the modularity function. Even when dealing with systems with strong spatial constraints, one might want to favor the importance of topological effects over spatial ones. A way of weighing both aspects in the modularity function is to introduce a mixing parameter, $\xi$, in order to interpolate between the two previous null models, $P_{ij}(\xi)=(\xi P^{\rm Spa}_{ij}+(1-\xi) P^{\rm NG}_{ij})$.

\section{Multi-scale Modularity}

Modularity optimization suffers from the limitation of producing one single partition, which is not satisfactory when dealing with multi-scale or hierarchical systems, that is systems made of (typically nested) modules at different scales. Different methods have been proposed to overcome this limitation \cite{lambi}. A first naive approach consists in re-applying modularity optimization on the communities found in the whole system. This approach provides a first guess but has the drawbacks of neglecting the global organization of the system when uncovering finer modules and of being unable to uncover coarser partitions than those obtained by the original modularity optimization. A second set of methods looks for local maxima of the modularity landscape, which has been shown to produce to modules at different scales \cite{Amaral07_2}. Finally, a third class of methods is based on multi-scale quality functions where a resolution parameter is incorporated such as to tune the characteristic size of the modules. A popular quantity is the parametric modularity introduced by Reichardt and Bornholdt \cite{reichardt}
\begin{equation}
\label{reichardt}
Q_\gamma = {1\over2m} \sum_{C \in \mathcal{P}} \sum_{i,j \in C} \biggl[ A_{ij} - \gamma P_{ij} \biggr],
\end{equation}
where $\gamma$ plays the role of a resolution parameter. Increasing $\gamma$ tends to decrease the characteristic size of the modules in the optimal partition.

\begin{figure}[!h]
    \centering
    \includegraphics[width=8cm,angle=0]{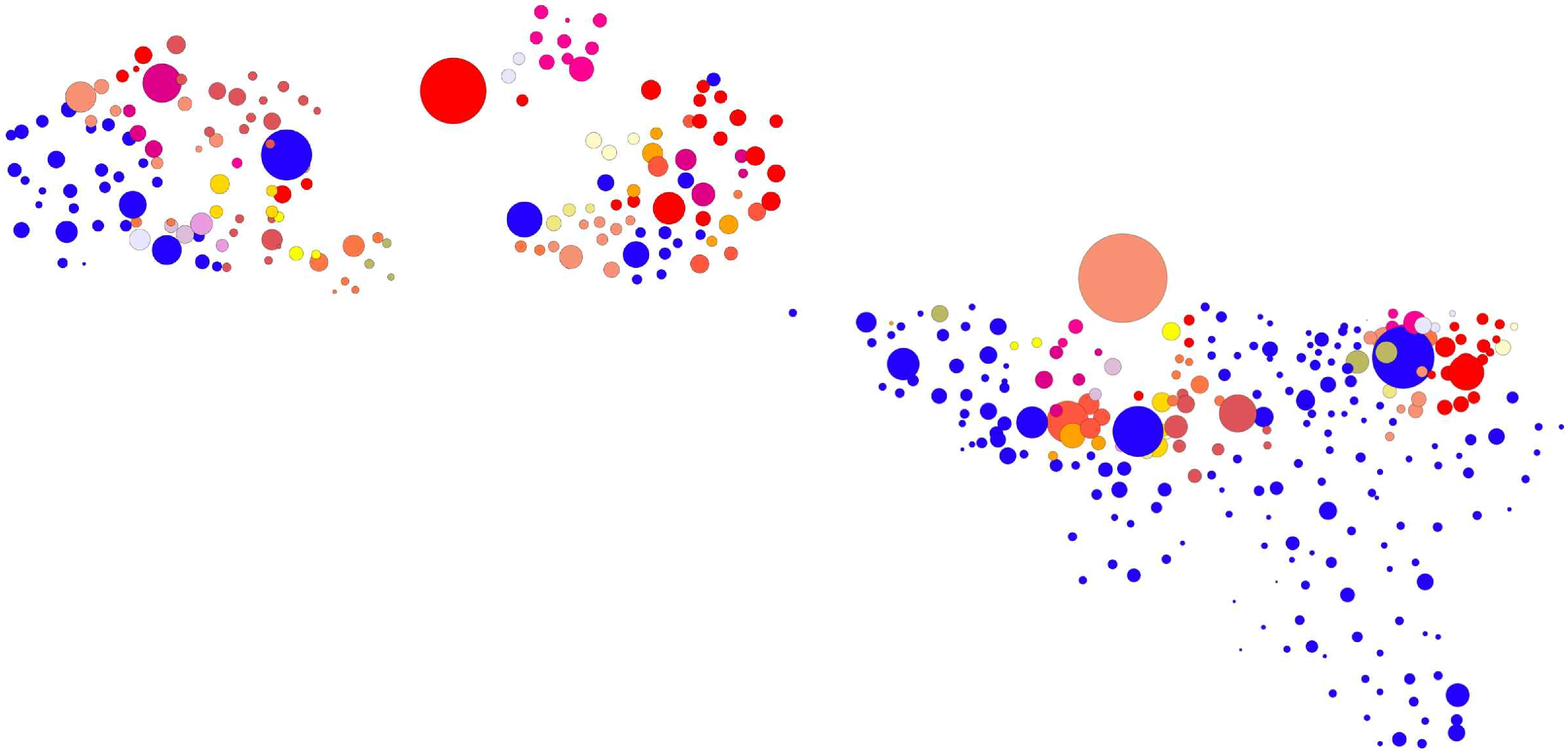}
    \caption{ Decomposition of the two main communities found in the Belgian mobile phone network into sub-communities (see main text). Each node represents a commune and its size is proportional to its number of clients $N_i$. (Left figure) Northern community. (Right figure) Southern community.\label{Hierarchy}}
\end{figure}

Because $Q_{\rm Spa}$ only differs from standard modularity by the choice of its null model, all three approaches can directly be applied  to search for multi-scale modules in spatial networks. While an exhaustive analysis of such hierarchical organization is beyond the scope of the
current work, we have performed a preliminary re-decomposition of the two largest communities found in the mobile phone network (see Fig.\ref{Hierarchy}). On finds $Q=0.019$ and $z$ score$=91$ in the case of the Northern community, and $Q=0.064$ and $z$ score$=425$ in the case of the Southern community (the $z$ score is calculated for the ensemble of random networks where weights are randomized). Values of the $z$ score are high but smaller than in the whole system ($z$ score$=803$). Moreover, the higher value observed in the Southern community than in the Northern community is expected due to the presence of bilingual Brussels in the former community. These results confirm that the linguistic division is the dominant factor, but also suggest that more regional factors (e.g. importance of local dialects, for instance in the Flemish community, cultural differences between cosmopolitan Brussels and the more rural Walloon, etc.) still play a role and lead to observable, finer sub-divisions of the country.

\section{Weights in the Belgian mobile phone data}
\begin{figure}[!h]
    \centering
    \includegraphics[width=8cm,height=6cm,angle=0]{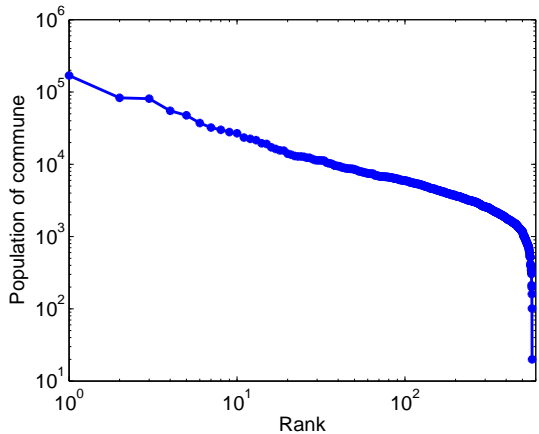}
    \caption{Zipf plot of the commune sizes. The system is highly heterogeneous with several orders of magnitude between largest and smallest communes.\label{ComZipf}}
\end{figure}

In our analysis of the mobile phone data, we have considered the fully connected matrix $\{A_{ij}\}_{i,j=1}^{571}$ where $A_{ij}$ is the total number of calls between users in commune $i$ and in commune $j$. Different types of weights could have been chosen for this aggregated network where node correspond to communes instead of individual users. In \cite{gauthier}, the authors  focus on another network where weights $A_{ij}/(N_{i}N_{j})$ correspond to the probability that users in $i$ and $j$ have called each other. This sensible choice gives, on average, the same importance to each commune and thus removes the effect of heterogeneity coming from different sizes of communes. In this article, we have instead preferred the first option, mainly for 2 reasons:

- i) One of the aims of modularity is to properly account for the importance of nodes in the null model, thereby producing balanced modules in terms of this measure of importance \cite{delvenne}. Because the definition of a proper null model is at the heart of this paper, we have preferred to preserve a strong heterogeneity (Fig. {\ref{ComZipf}}) in the system and to let the definition of modularity ``deal with it".

-ii) By focusing on a meta-network where the weights of the links between communes is the sum over the links between their users, the same importance is given to each user. More importantly, modularity at the commune-level is related to modularity at the user-level: by optimizing the modularity of $A_{ij}$, one finds the best partition of the user network with the constraint that users in the same commune must be in the same community \cite{arenas}.

\begin{figure}[!h]
    \centering
    \includegraphics[width=8cm,height=6cm,angle=0]{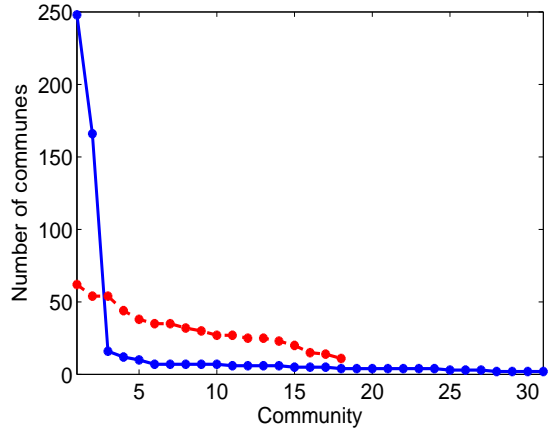}
    \caption{Sizes of the communities found by Spa (full blue line) and NG (dashed red line). The size of each community is measured by the number of communes it contains. In the partition found by Spa, two communities are large while the others are of negligible size. In the partition found by NG, all communities are of similar size.\label{ComSize}}
\end{figure}

\begin{figure}[!h]
    \centering
    \includegraphics[width=8cm,height=6cm,angle=0]{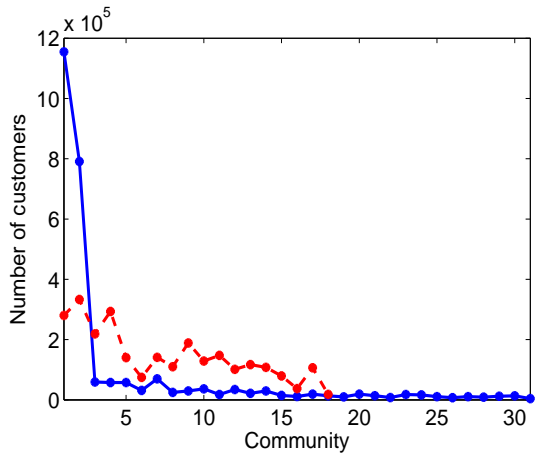}
    \caption{Sizes of the communities found by Spa (full blue line) and NG (dashed red line). The size of each community is measured by the number of customers living in it. The labelling of the communes is the same as in the previous figure. The partition found by Spa also gives two large communities while the others are of negligible size. In the partition found by NG, all communities are again of similar size.\label{ComSizeCust}}
\end{figure}

\section{Size of the communities}
In the main text, we point out that the two largest communities found using the spatial null-model account for about 75\% of all communes. The remaining communes are assigned to 29 small communities, most of them close to Brussel. This can be attributed to the blindness of the algorithm we used to overlapping communities and the strong interaction of Flemish speaking communes with Brussel. To clearly illustrate this point, we plot in Fig. \ref{ComSize} the size of the communities found by the two null-models. This plot clearly shows that the communities found by Spa other than the two largest are of negligible size. The sizes of the communities found by NG on the other hand are rather homogeneous. Fig. \ref{ComSizeCust} shows the size of the communities in term of number of customer. Again the two largest communities found by Spa account for more than 70\% of the customers, while NG again divides the Belgian population into communities fairly similar in size.

\begin{figure}[!h]
    \centering
    \includegraphics[width=8cm,height=6cm,angle=0]{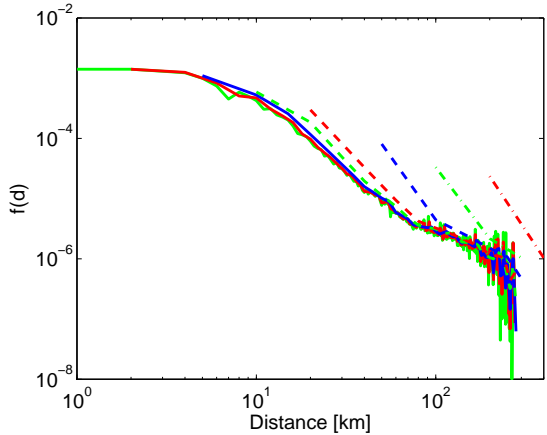}
    \caption{Deterrence function $f(d)$ for different size of bins. Solid lines: green: 1 [km], red: 2 [km], blue: 5 [km]. Dashed lines: green: 10 [km], red: 20 [km], blue: 50 [km]. Dashed and dotted line: green: 100 [km], red: 200 [km].\label{Bins_f_d}}
\end{figure}

\begin{figure}[!h]
    \centering
    \includegraphics[width=8cm,height=6cm,angle=0]{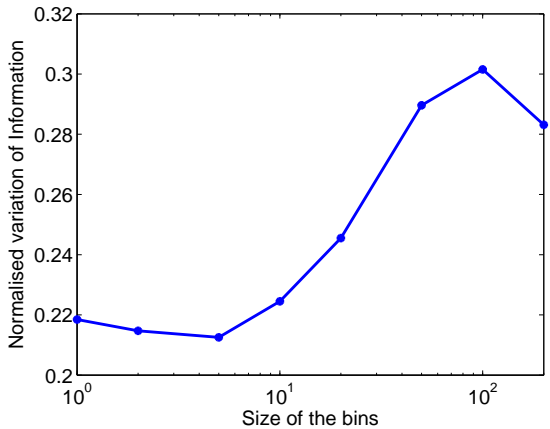}
    \caption{Averaged normalised variation of information as a function of bin size. The minimum is reached at 5 [km].\label{Bins_VI}}
\end{figure}

\begin{figure}[!h]
    \centering
    \includegraphics[width=8cm,height=6cm,angle=0]{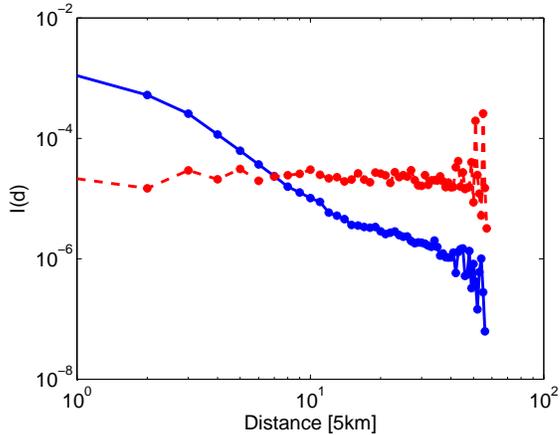}
    \caption{Solid (blue) line: $f(d)$ for the Belgian mobile phone network, dashed (red) line: $f(d)$ after the positions' randomisation. \label{polo}}
\end{figure}

\section{Binning distance}
The evaluation of $f(d)$ depends on the size of the bins used to measure distances. Two extreme cases are 1 [km] and 200 [km] (the largest distances in Belgium are of the order of 300 [km] and we need at least two bins). To choose the most appropriate size for the bins in that range, we computed the deterrence function $f(d)$ and the partitions obtained for 8 different bin sizes $s$: 1, 2, 5, 10, 20, 50, 100 and 200 [km]. The different deterrence functions are shown in Fig. \ref{Bins_f_d}. There is no clear discrimination for distances smaller than 5 [km] and the noise in the tail of the distribution is negligible from 20 [km]. Considering the size of Belgium, the number of communes (571) and the typical distance between them  a bin size of 5 [km] is a reasonable choice \cite{mobile}.

In order to support this choice, we have also computed the average normalised variation of information  $\langle V(s) \rangle \equiv \frac{1}{N_s} \sum_{s^{'} \neq s} V(s,s^{'})$ between the partition at bin size $s$ and those at other bin sizes (see Fig. \ref{Bins_VI}). The size that is closest to all the others, thus the most representative of the system, is 5 [km].

In Figure \ref{polo}, we plot $f(d)$ when measure with bin sizes of 5[km] in the original data and in a the same network where positions are randomized.

\section{Gravity Model Benchmark: Averages}
\begin{figure}[!h]
    \centering
    \includegraphics[width=6cm,angle=0]{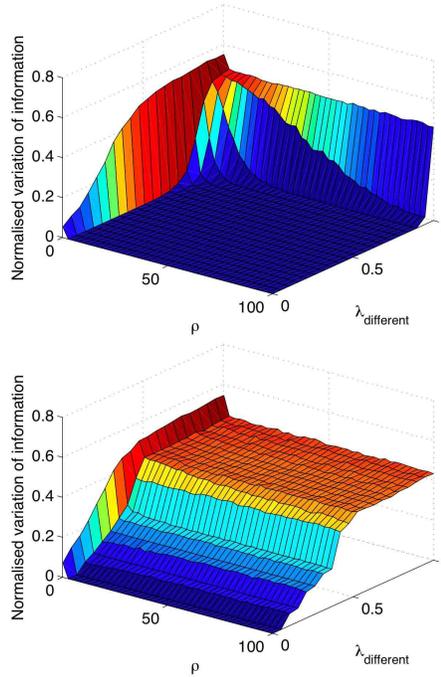}
    \caption{Averaged normalised variation of information over the $(\lambda_{\rm different},\rho)$ parameter space for Spa (upper figure) and NG (lower figure) when tested on the spatial benchmark. Spa is able to recover the correct communities over a wide range of parameters' values, while GN fails to find the correct communities almost as soon as the interaction $\lambda_{\rm different}$ is turned on.\label{VIBench_average}}
\end{figure}

\begin{figure}[!h]
    \centering
    \includegraphics[width=8cm,height=6cm,angle=0]{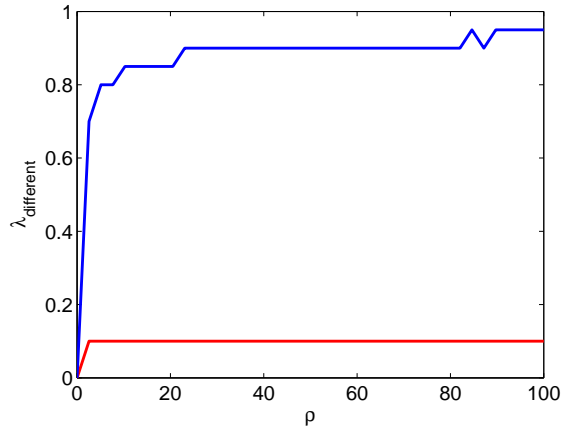}
    \caption{$\lambda_{\rm different}-\rho$ parameter space separation. Solid red line: NG, solid blue line: Spa. Below the line, the partition found matches the real one, above there are discrepancies. Spa is able to recover the original communities on a much larger range of parameters\label{Bench_Sep}}
\end{figure}

\begin{figure}[!h]
    \centering
    \includegraphics[width=8cm,height=6cm,angle=0]{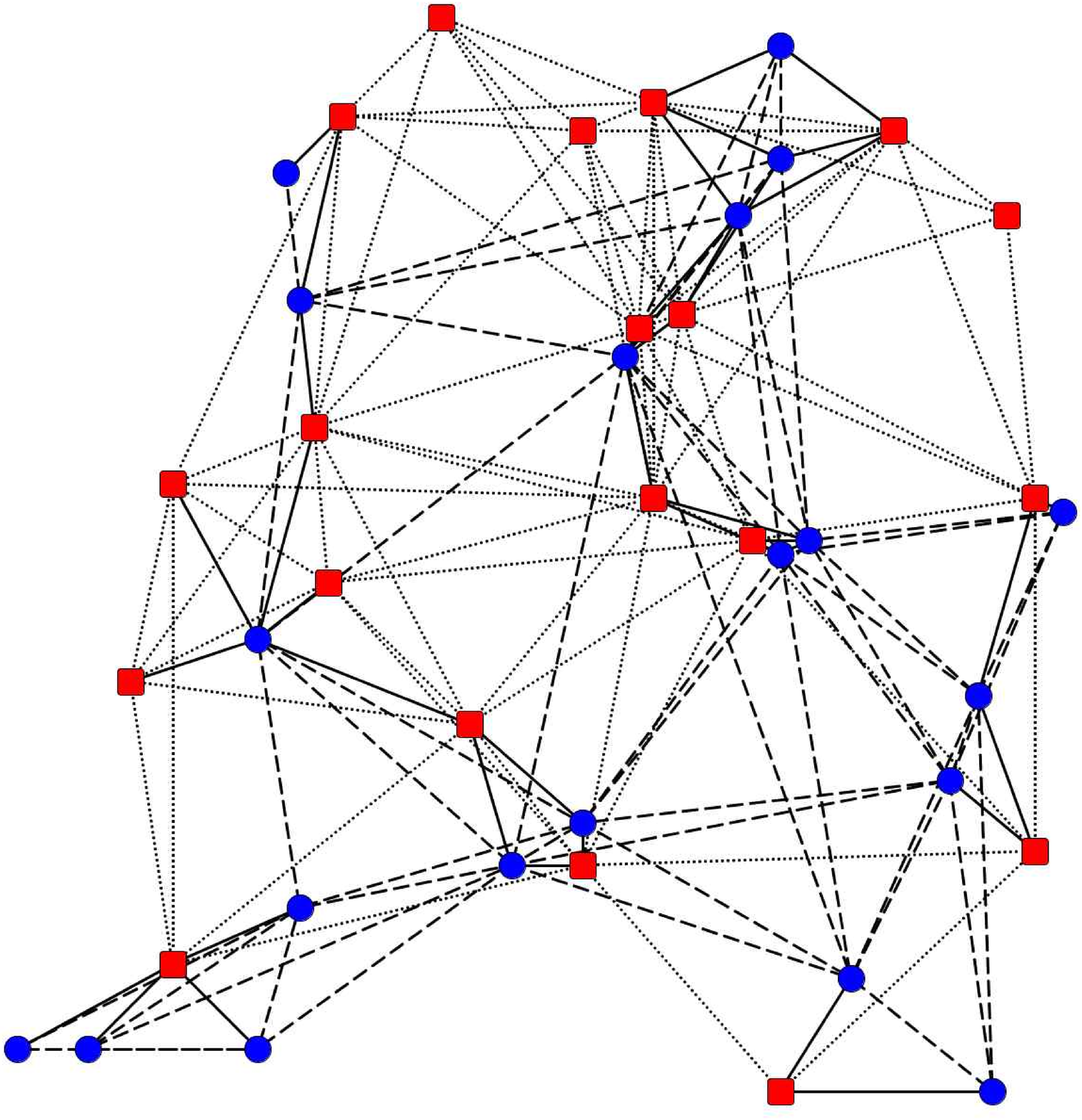}
    \caption{Illustration of the probabilities used for a Gravity Model Benchmark made of 40 nodes with $\lambda_{\rm different}=0.5$. Only $20 \%$ of links with highest probability are represented. Hidden communities are represented by squares and circles. Different types of interactions are highlighted by their linestyle: red-red: dotted, blue-blue: dashed, red-blue: solid. \label{visu}}
\end{figure}

In the main text, we tested how close uncovered partitions were from the known underlying community structure as a function of the interaction strength and the density of links. The results for one single realization of the benchmarks were overwhelmingly in favour of Spa. Here we produce the same graphs,  but averaged over 100 different realisations of the random networks, thus leading to a smoother surface, Fig. \ref{VIBench_average}. We also present a ``phase diagram" ($\lambda_{\rm different},\rho$) in which we highlight values where the partitioning ceases to be perfect (i.e. the normalised variation of information becomes larger than 0), Fig. \ref{Bench_Sep}. One observes that Spa offers a perfect reconstruction over a significantly broader range of parameters than NG. N.B. Visualizations of the benchmarks are shown in Fig. \ref{visu}.

\section{Bipartition of the mobile phone data}
\begin{figure}[!h]
    \centering
    \includegraphics[width=8cm,height=3.8cm,angle=0]{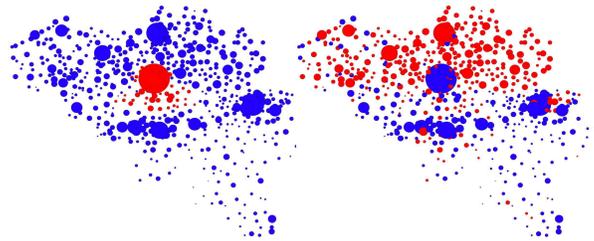}
    \caption{ Decomposition of a Belgian mobile phone network into communities (see main text). Each node represents a commune and its size is proportional to its number of clients $N_i$. (Left figure) Partition into 18 communities found by optimizing GN modularity. (Right figure) Partition into 31 communities found by optimizing  Spa modularity.\label{Bipart}}
\end{figure}

From modularity, it is always possible to partition a network into two communities by assigning each node to a community according to its sign in the leading eigenvector of the modularity matrix. In this procedure, a negative second  largest eigenvalue indicates that this bipartition is a good approximation to the full optimisation of modularity \cite{newman}. When applied to the Belgian mobile phone network, the second eigenvalue for NG modularity matrix is positive, contrary to Spa, thereby suggesting that a bipartition is a reasonable approximation to the full optimization for Spa. This is confirmed visually in Fig. \ref{Bipart}, where NG picks Brussels and its neighborhood, and the rest of Belgium as the best bipartition, while Spa gives a North-South bipartition consistent with the linguistic bipartition of the country.

\begin{thebibliography}{}


\bibitem{review2}
M.E.J. Newman, A.-L. Barab{\'a}si and D.J. Watts, {\it The Structure and Dynamics of Networks} (Princeton University Press, Princeton, 2006).

\bibitem{bart_review} M. Barthelemy (2010) Spatial Networks, arXiv:1010.0302

\bibitem{gast1}
M. T. Gastner  and M. E. J. Newman (2006) The spatial structure of networks, {\em The Eur. Phys. J. B} {\bf 49}, pp. 247-252.

\bibitem{cost1}
R. Kitchin and M. Dodge (2000) ÔPlacingÕ Cyberspace: Geography, Community And Identity, {\em Information Technology, Education and Society} {\bf 1}, pp. 25-46.
\bibitem{Amaral00}
LAN. Amaral, A. Scala, M. Barthelemy, HE. Stanley (2000) Classes of small-world networks, {\em Proc. Natl. Acad. Sci. USA} {\bf 97}, 11149-11152
\bibitem{Amaral04}
R. Guimera, LAN. Amaral (2004) Modeling the world-wide airport network, {\emph Eur. Phys. J. B} {\bf 38}, 381-385
\bibitem{cost2}
Kaiser M, Hilgetag CC (2004) Modelling the development of cortical systems networks, {\em Neurocomputing} {\bf 58}, pp. 297Ð302.
\bibitem{cost2b}
A. Barrat, M. BarthŽlemy and A. Vespignani (2005) The effects of spatial constraints in the evolution of weighted complex networks, {\em J. Stat. Mech.}, P05003.
\bibitem{cost3}
Y. Hayashi (2006) A Review of Recent Studies of Geographical Scale-Free Networks, {\em IPSJ Journal} {\bf 47}, 776.
\bibitem{cost4}
J.N. Cummings and S. Kiesler (2007) Coordination costs and project outcomes in multi-university collaborations, {\em Research Policy} {\bf 36}, pp. 138-152.
\bibitem{liben}
D. Liben-Nowell, J. Novak, R. Kumar, P. Raghavan and A. Tomkins (2005) Geographic routing in social networks, {\em Proc. Natl. Acad. Sci. USA} {\bf 102}, 11623-11628
\bibitem{cost5}
D. Bassett, E. Bullmore, A. Meyer-Lindenberg, J. Apud, D. Weinberger and R. Coppola (2009) Cognitive fitness of cost-efficient brain functional networks, {\em Proceeding of the National Academy of Sicences} {\bf 106}, 11747.

\bibitem{gr1}
V. Carrothers (1956) A Historical Review of the Gravity and Potential Concepts of Human Interaction, {\em J. of the Am. Inst. of Planners} {\bf 22}, pp. 94-102.
\bibitem{gr3}
A.G. Wilson (1967) A statistical theory of spatial distribution models, {\it Transportation Research} {\bf 1}, pp. 253-269.
{\bf 54}, pp. 68-78
\bibitem{gr5}
S. Erlander and N. F. Stewart (1990) The Gravity Model in Transportation Analysis: Theory and Extensions, Brill Academic Publishers, Utrecht.
\bibitem{gravity1}
K. Bhattacharya, G. Mukherjee, J. Saram\"aki, K. Kaski and S.S. Manna (2008) The International Trade Network: weighted network analysis and modelling, {\em J. Stat. Mech.}, P02002.
\bibitem{gravity1b}
M. Levy (2010) Scale-free human migration and the geography of social networks, {\em Physica A} {\bf 389}, 4913-4917.
\bibitem{gravity2}
W.-S. Jung, F. Wang and H.E. Stanley (2008) Gravity model in the Korean highway, {\em Europhys. Lett.} {\bf 81}, 48005.
\bibitem{mobile}
R. Lambiotte, V.D. Blondel, C. de Kerchove, E. Huens, C. Prieur, Z. Smoreda and P. Van Dooren (2008) Geographical dispersal of mobile communication networks, {\it Physica A} {\bf 387}, 5317-5325.
\bibitem{mobile2}
G.M. Krings, F. Calabrese, C. Ratti, V.D. Blondel (2009) Urban gravity: a model for inter-city telecommunication flows, {\em J. Stat. Mech.}, L07003.

\bibitem{kleinberg}
J. Kleinberg (2000) Navigation in a small world, {\em Nature} {\bf 406}, 845.
\bibitem{bart}
M. Barth\'elemy (2003) Crossover from scale-free to spatial networks, {\em Europhys. Lett.} {\bf 63}, 915.
 \bibitem{Manna}
S.S. Manna and P. Sen (2002) Modulated scale-free network in Euclidean space, {\em Phys. Rev. E} {\bf 66}, 066114.
 \bibitem{gast2}
M.T. Gastner and M.E.J. Newman (2006) Optimal design of spatial distribution network, {\em Phys. Rev. E} {\bf 74}, 016117.
 \bibitem{wong}
L.H. Wong, P. Pattison and G. Robins (2006) A spatial model for social networks, {\em Physica A} {\bf 360}, pp. 99-120.
\bibitem{kosmidis}
K. Kosmidis, S. Havlin and A. Bunde (2008) Structural properties of spatially embedded networks, {\em Europhys. Lett.} {\bf 82}, 48005.
\bibitem{mascolo}
S. Scellato, C. Mascolo, M. Musolesi and V. Latora (2010) Distance Matters: Geo-social Metrics for Online Social Networks. {\em Proceedings of WOSNÕ 10}.
\bibitem{colizza}
V. Colizza, A. Barrat, M. Barthelemy and A. Vespignani (2006) The role of the airline transportation network in the prediction and predictability of global epidemics, {\em Proc. Natl. Acad. Sci. USA} {\bf 103}, 2015.


\bibitem{GN}
M. Girvan and M. E. J. Newman (2002) Community structure in social and biological networks, {\em Proc. Natl. Acad. Sci. USA} {\bf 99}, pp. 7821Ð7826.
\bibitem{danon}
L. Danon, A. D\'iaz-Guilera, J. Duch and A. Arenas (2005) Comparing community structure identification, {\em J. Stat. Mech.} P09008.
\bibitem{newman_modul_PNAS}
M.E.J. Newman (2006) Modularity and community structure in networks, {\it Proc. Natl. Acad. Sci. USA} {\bf 103}, pp. 8577-8582.
\bibitem{Amaral05a}
R. Guimera and LAN. Amaral (2005) Functional cartography of complex metabolic networks, {\emph Nature} {\bf 433}, 895-900
\bibitem{sune}
N. Gulbahce and S. Lehmann (2008) The art of community detection, {\em BioEssays} {\bf 30}, pp. 934-938.
\bibitem{porter}
M.A. Porter, J.-P. Onnela and P.J. Mucha (2009) Communities in Networks, {\em Notices of the American Mathematical Society} {\bf 56}, pp. 1082-1097, 1164-1166.
\bibitem{fort}
S. Fortunato (2010) Community detection in graphs, {\em Physics Reports} {\bf 486}, pp. 75-174.

\bibitem{brockmann10}
D. Brockmann (2010) Following the Money Physics World, February, 31.
\bibitem{brockmann10b}
C. Thiemann, F. Theis, D. Grady, R. Brune, D. Brockmann (2010) The structure of borders in a small world, {\em PLoS ONE} {\bf 5}, e15422.
\bibitem{gauthier}
V.D. Blondel, G. Krings and I. Thomas (2010) Regions and borders of mobile telephony in Belgium and in the Brussels metropolitan zone, {\em Brussels Studies} {\bf 42}, 4.
\bibitem{ratti}
C. Ratti, S. Sobolevsky, F. Calabrese, C. Andris, J. Reades, M. Martino, R. Claxton, S.H. Strogatz (2010) Redrawing the Map of Great Britain from a Network of Human Interactions, {\em PLoS ONE} {\bf 5}, e14248.

\bibitem{RW87}
  T.E. Rihll, A.G. Wilson (1987)
  Spatial Interaction and Structural Models in Historical Analysis: Some Possibilities and an Example
  \emph{Histoire \& Mesure} \textbf{2} pp. 5-32.

\bibitem{KER08}
  C. Knappett, T.S. Evans, R.J. Rivers (2008)
  Modelling Maritime Interaction In The Aegean Bronze Age
  \emph{Antiquity} \textbf{82} pp. 1009-1024.

\bibitem{bassett}
D. Bassett, D. Greenfield, A. Meyer-Lindenberg, D. Weinberger, S. Moore and E.T. Bullmore (2010) Efficient physical embedding of topologically complex information processing networks in brains and computer circuits, {\em PLoS Computational Biology} {\bf 6}, e1000748.
\bibitem{frontiers}
D. Meunier, R. Lambiotte and E.T. Bullmore (2010) Modular and hierarchically modular organization of brain networks, {\em Frontiers in NeuroScience} {\bf 4}, 200.

\bibitem{Guimera}
R. Guimer\`a, S. Mossa, A. Turtschi and L.A.N. Amaral (2005) The worldwide air transportation network: anomalous centrality, community structure, and cities' global roles, {\em Proc. Natl. Acad. Sci. USA} {\bf 102}, 7794.
\bibitem{onnela2010}
J.-P. Onnela, S. Arbesman, A.-L. Barab\'asi and N.A. Christakis (2010) Geographic constraints on social network groups, {\em arxiv:1011.4859}.

\bibitem{davis70}
J.A. Davis (1970) Clustering and hierarchy in interpersonal relations: testing two graph theoretical models on 742 sociomatrices, {\em American Sociological Review} {\bf 35}, pp. 843-851.
\bibitem{gouldner60}
A.W. Gouldner (1960) The norm of reciprocity: a preliminary statement, {\em American Sociological Review} {\bf 25}, pp. 161-178.
\bibitem{podolny94}
J.M. Podolny (1994) Market uncertainty and the social character of economic exchange, {\em Administrative Science Quarterly} {\bf 39}, pp. 458-483.

\bibitem{mcpherson01}
J.M. McPherson, L. Smith-Lovin and J.M. Cook (2001) Birds of a feather: homophily in social networks, {\em Annual Review of Sociology} {\bf 27}, pp. 415-444.
\bibitem{feld81}
S.L. Feld (1981) The focused organization of social ties, {\em American Journal of Sociology} {\bf 86}, pp. 1015-1035.
\bibitem{kossinets06}
G. Kossinets and D,J, Watts (2006) Empirical analysis of an evolving social network, {\em Science} {\bf 311}, pp. 88-90.
\bibitem{owen04}
J. Owen-Smith and W.W. Powell (2004) Knowledge networks as channels and conduits: The effects of spillovers in the Boston biotechnology community, {\em Organization Science} {\bf 15}, pp. 5-21.
\bibitem{crandall}
D.J. Crandall, L. Backstrom, D. Cosley, S. Suri, D. Huttenlocher and J. Kleinberg (2010), Inferring social ties from geographic coincidences, PNAS Early Edition
\bibitem{schelling}
T.C. Schelling (1971) Dynamic Models of Segregation, {\em J. of Math. Soc.} {\bf 1}, pp. 143-186.
\bibitem{watts}
D.J. Watts, P.S. Dodds and M.E.J. Newman (2002) Identity and Search in Social Networks, {\em Science} {\bf 296}, pp. 1302-1305
\bibitem{pietro}
T.S. Evans, R. Lambiotte, P. Panzarasa (2010) Communities and Patterns of Scientific collaboration, arXiv:1006.1788


\bibitem{NG}
M.E.J. Newman and M. Girvan (2004) Finding and evaluating community structure in networks, {\it Phys. Rev. E} {\bf 69}, 026113.

\bibitem{opt2}
M. E. J. Newman (2006) Finding community structure in networks using the eigenvectors of matrices, {\em Phys. Rev. E} {\bf 74}, 036104.

\bibitem{Amaral07_1}
R. Guimera, M. Sales-Pardo, LAN. Amaral (2007) Module identification in bipartite and directed networks, {\emph Phys. Rev. E} {\bf 76}, 036102
\bibitem{gen1}
V. Nicosia, G. Mangioni, V. Carchiolo and M. Malgeri (2009) Extending the definition of modularity to directed graphs with overlapping communities, {\it J. Stat. Mech.}, P03024.
\bibitem{gen2}
M.J. Barber (2007) Modularity and community detection in bipartite networks, {\it Phys. Rev. E} {\bf 76}, 066102.

\bibitem{opt1}
R. Guimer\'a, M. Sales-Pardo, and L.A.N. Amaral (2004) Modularity from fluctuations in random graphs and complex networks, {\em Phys. Rev. E} {\bf 70}, 025101(R).
\bibitem{opt3}
V.D. Blondel, J.-L. Guillaume, R. Lambiotte and E. Lefebvre (2008) Fast unfolding of community hierarchies in large networks, {\em J. Stat. Mech.}, P10008.
\bibitem{spectmas}
T. Richardson, P.J. Mucha and M.A. Porter (2009) Spectral tripartitioning of networks, {\em Phys. Rev. E.} {\bf 80}, 036111.

\bibitem{FB}
S. Fortunato and M. Barth{\'e}lemy (2007) Resolution limit in community detection, {\it Proc. Natl. Acad. Sci. USA} {\bf 104}, pp. 36-41.
\bibitem{Amaral07_2}
M. Sales-Pardo, R. Guimera, AA. Moreira, LAN. Amaral (2007) Extracting the hierarchical organization of complex systems, {\emph Proc. Natl. Acad. Sci. USA} {\bf 104}, 15224-15229
\bibitem{Good}
 B.H. Good, Y.-A. de Montjoye and A. Clauset (2010) The performance of modularity maximization in practical contexts, {\em Phys. Rev. E} {\bf 81}, 046106.
 \bibitem{reichardt}
J. Reichardt and S. Bornholdt (2006) Statistical Mechanics of Community Detection, {\it Phys. Rev. E} {\bf 74}, 016110.

\bibitem{varinfo}
M. Meila (2003) Comparing Clusterings by the Variation of Information, {\em Learning Theory and Kernel Machines}, 173-187.

\bibitem{cairncross97}
F. Cairncross (1997) The Death of Distance, Harvard University Press, Cambridge MA.

\bibitem{spiky}
R, Florida (2005) The World Is Spiky, The Atlantic Monthly, October, 48

\bibitem{mobility}
C. Song, Z. Qu, N. Blumm and A.-L. Barab\'asi (2010) Limits of Predictability in Human Mobility, {\em Science} {\bf 327}, 1018.

\bibitem{ABL09}
Y.-Y. Ahn, J.P. Bagrow and S. Lehmann (2010) Link communities reveal multi-scale complexity in networks, {\em Nature} {\bf 466} , 761Ð764.
\bibitem{EL10}
T.S. Evans and R. Lambiotte (2010) Edge Partitions and Overlapping Communities in Complex Networks, {\it Eur. Phys. J. B} {\bf 77}, 265-272.



 \bibitem{lambi}
R. Lambiotte (2010) Multi-scale Modularity in Complex Networks, Modeling and Optimization in Mobile, Ad Hoc and Wireless Networks (WiOpt), 2010 Proceedings of the 8th International Symposium on, 546-553.




\bibitem{delvenne}
J.-C. Delvenne, S. N. Yaliraki, and M. Barahona, (2010), Stability of graph communities across time scales, \textit{PNAS}  \textbf{107} (29) 12755-12760

\bibitem{arenas}
Arenas A, Duch J, Fern\'andez A and G\'omez S, (2007) {\em N. J. of Phys.} {\bf 9} 176.


\bibitem{newman}
M.E.J. Newman, (2006), Finding community structure in networks using the eigenvectors of matrices, \textit{Phys. Rev. E}, \textbf{74}, 036104,



\end{thebibliography}
\end{document}